\documentclass[12pt,a4paper]{article}

\usepackage[bf]{caption}
\usepackage{graphicx}

\makeatletter
\renewcommand\section{\suppressfloats[t]%
                      \@startsection{section}{1}{\z@}%
                                    {3.25ex \@plus1ex \@minus.2ex}%
                                    {-1em}%
                                    {\normalfont\normalsize\bfseries}}
\makeatother

\begin{document}

\sffamily

\begin{center}
  {\large               
    {\bfseries  Atomic Self-Diffusion in Quasicrystals:\\ A Molecular
    Dynamics 
    Study 
    }\\[4mm]
        Johannes Roth$^{1}$, Franz G\"ahler$^{2}$
  }\\[2mm]
  $^{1}$ Institut f\"ur Theoretische und Angewandte Physik,
  Universit\"at Stuttgart, 70550 Stuttgart, Germany\\[2mm]
  $^{2}$ Centre de Physique Th\'eorique, Ecole
  Polytechnique, 91128 Palaiseau, France
\end{center}
\vspace*{1mm}

\rmfamily

\section*{Abstract}
We present a molecular dynamics study on atomic self-diffusion in Frank-Kasper
type dodecagonal quasicrystals. It is found that the
quasicrystal-specific flip mechanism for atomic diffusion as predicted 
by Kalugin and Katz\cite{kuk}, indeed occurs in this system. However, in order 
to be effective, this mechanism needs to be catalyzed by other defects
such as half-vacancies if the structure is truly three-dimensional. 
For this reason, flip diffusion is difficult to distinguish from
standard vacancy diffusion. In a quasi-two-dimensional setup, however,
the flips may occur without other defects. Activation energies and
flip frequencies are also determined.

\section{Introduction}

Quasicrystals are by now well-established in solid state physics. They
exhibit non-crystallographic symmetries together with a number of
unusual physical properties. Recent reviews may be found in
Refs.~\cite{hip}. 

There has recently been much interest in atomic self-diffusion in
quasicrystals, mostly
triggered by a paper by Kalugin and Katz\cite{kuk} 
where a diffusion mechanism specific to quasicrystals was proposed.
The elementary process in this {\it flip mechanism} 
consists of certain quasicrystal-specific rearrangements of atoms, 
where the initial and final configurations are energetically almost 
degenerate. In quasicrystals which are decorations of quasiperiodic
tilings, the flip mechanism consists of a reshuffling of certain local
tile configurations, along with their decorations \cite{kuk}.

Up to now experiments have not given a clear answer to the diffusion
behaviour of quasicrystals. Coddens and his group \cite{cod} found
atomic hopping of Cu and Fe in AlCuFe quasicrystals. The process can
be regarded as flips but it behaves differently from the proposal of Kalugin
and Katz. 
Tracer diffusion has been studied by Nakajima et
al.~\cite{nak}. Their results indicate an unusually low diffusivity of Mn
in AlPdMn quasicrystals as compared to crystals of similar composition. 
To shed more light on these results one has to use
simulations to find the elementary diffusion moves and their
energetical properties.

Flip diffusion has so far been studied in pure tiling models, without 
bothering about a specific atomic decoration of the tilings \cite{jos1}. 
While such an approach indeed proves that elementary flip processes 
do add up to diffusive behaviour, the physical feasibility of the
flip mechanism and the magnitude of flip diffusion remain much less
certain. In particular, activation energies of elementary flips cannot
be estimated without a concrete atomic structure, and in a
reshuffling of tiles atoms may have to move only much smaller
distances than the vertices of the tiles. Those questions cannot be 
studied without having a specific atomic structure in mind.
Some preliminary work treating this problem has been
published in Ref.~\cite{gae3}. 

We propose to study the feasibility of flip diffusion in a 
concrete atomic model quasicrystal, by means of molecular dynamics (MD)
simulations. For such a simulation, not only a realistic model structure,
but also (short range) interatomic potentials stabilizing the
model structure are needed. Fortunately, this has become available:
Dzugutov \cite{dzu} has found a one-component system with a simple potential,
which solidifies into a quasicrystalline structure known already
before as a realistic model of dodecagonal quasicrystals \cite{gae2,bee}.

In Section 2, we shall describe this model structure in more detail,
along with some ideas on potential flip moves, before we present our 
MD simulation results in Section 3. In Section 4 activation energies
and frequencies for the flips are given. We finally conclude in Section 5.

\section{Description of the Model Structures}

\begin{figure}[ht!]
\centerline{\includegraphics[width=14cm]{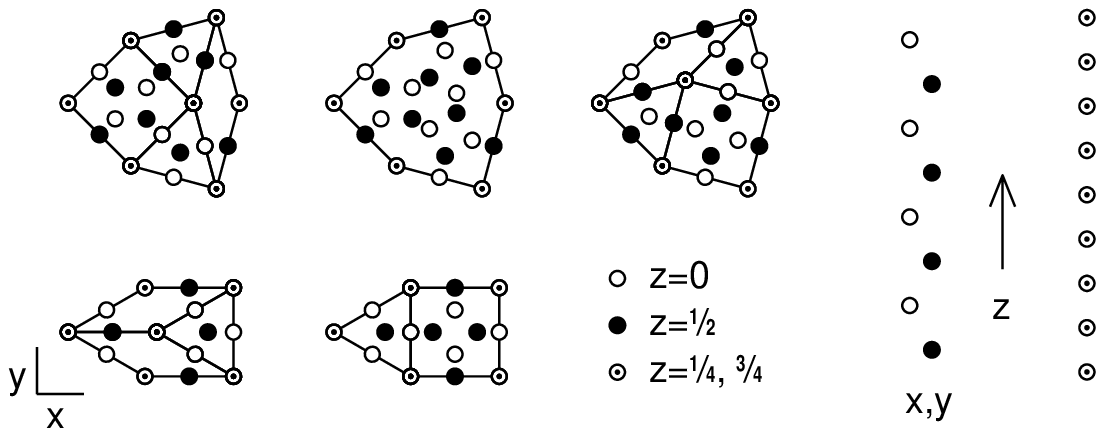}}

\caption{\small 
   The basic tiles with their decorations. Height coordinates are
   in fractions of the period length. Two elementary rearrangements 
   of tiles and atoms are shown. One subdivision of a hexagon is 
   replaced by a different one through an intermediate step (top line),
   and two rhombi are replaced by a square (bottom line). In all these 
   elementary steps, a whole column of (dotted) $A$ layer atoms 
   is replaced by a slightly jagged column of $B$/$\bar{B}$ 
   layer atoms, or vice versa. These columns are shown on the right.
   A whole column thus has to move a quarter of a period length upwards
   or downwards.
   }
\label{basictile}
\end{figure}

The structure found by Dzugutov \cite{dzu} in his MD simulation is a
layered structure which, apart from some defects, is periodic in 
one direction, but quasiperiodic and 12-fold symmetric in the 
plane perpendicular to it. It basically is of Frank-Kasper type,
i.e., it is mostly tetrahedrally close-packed, and can be described
as a periodic stacking $ABA\bar B$ of a dodecagonal 
layer $A$ and two hexagonal layers, $B$ and 
$\bar B$, which are just rotated by $30^\circ$ with respect to each other.
The atoms in layer $A$ form the vertices of a simple tiling of squares,
triangles, asymmetric hexagons and $30^\circ$~rhombi.
The whole structure is, in fact, a decoration of such a
tiling \cite{gae2,bee}. The tiles occurring in the structure, together with 
their decorations, are shown in Fig.~\ref{basictile}.

All structures built on a tiling with squares and triangles only are 
perfectly tetrahedrally close-packed. Such structures are therefore
very rigid, and there are also very few small groups of tiles that
can be reshuffled. If hexagons or rhombi are present, 
however, some atoms are 
not close-packed: there are some octahedra occurring in the interior 
of hexagons, and near the obtuse corners of rhombi. Near those
octahedra, the structure is much softer.
Moreover, in such structures
there are also many local tile configurations which can easily be 
reshuffled. Some of these tile flips are shown in Fig.~\ref{basictile}.
In all the  
moves, a whole column of atoms has to move upwards or downwards
(Fig.~\ref{basictile}), 
whereas other atoms move very little. Obviously, such a move is very 
unlikely, since the vertical distance of atoms is already rather small. 
If one introduces a vacancy in the column, however, the move becomes 
much easier. Since between two atom positions in a column there is
another good atom position, a neighboring atom can then move half-way
into the vacancy, effectively splitting it into two half-vacancies,
which then can move up and down the column, thereby transforming it. 
Such half-vacancies therefore efficiently catalyze the moves shown in 
Fig.~\ref{basictile}. They are responsible for the breaking of
periodicity in z-direction. 
Note that the movement of atoms in the xy-plane is much smaller than
that of the vertices.
In our simulations, structures built on several different tilings are
used, both perfectly ordered as well as disordered ones. Two different
geometries have been studied: a cubic box with each sample containing about
8000~atoms and a flat tetragonal box with aspect ratio of about 7.5:1
containing 14040~atoms.

\section{Results of the Molecular Dynamics Simulations}

\begin{figure}[ht!]
\centerline{\parbox{7.0cm}{\includegraphics[width=7cm]{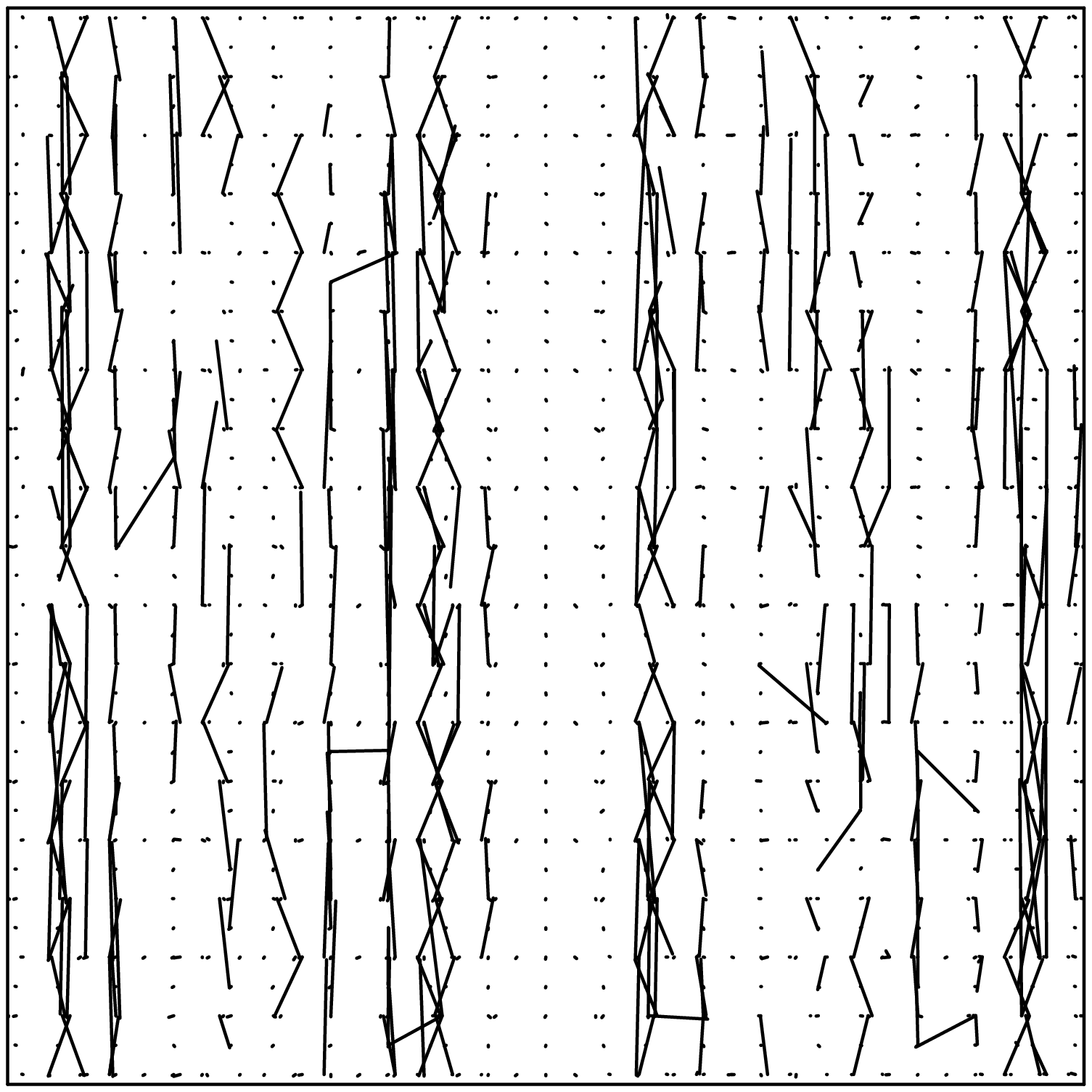}}
\hspace*{1cm}\parbox{7.0cm}{\includegraphics[width=7cm]{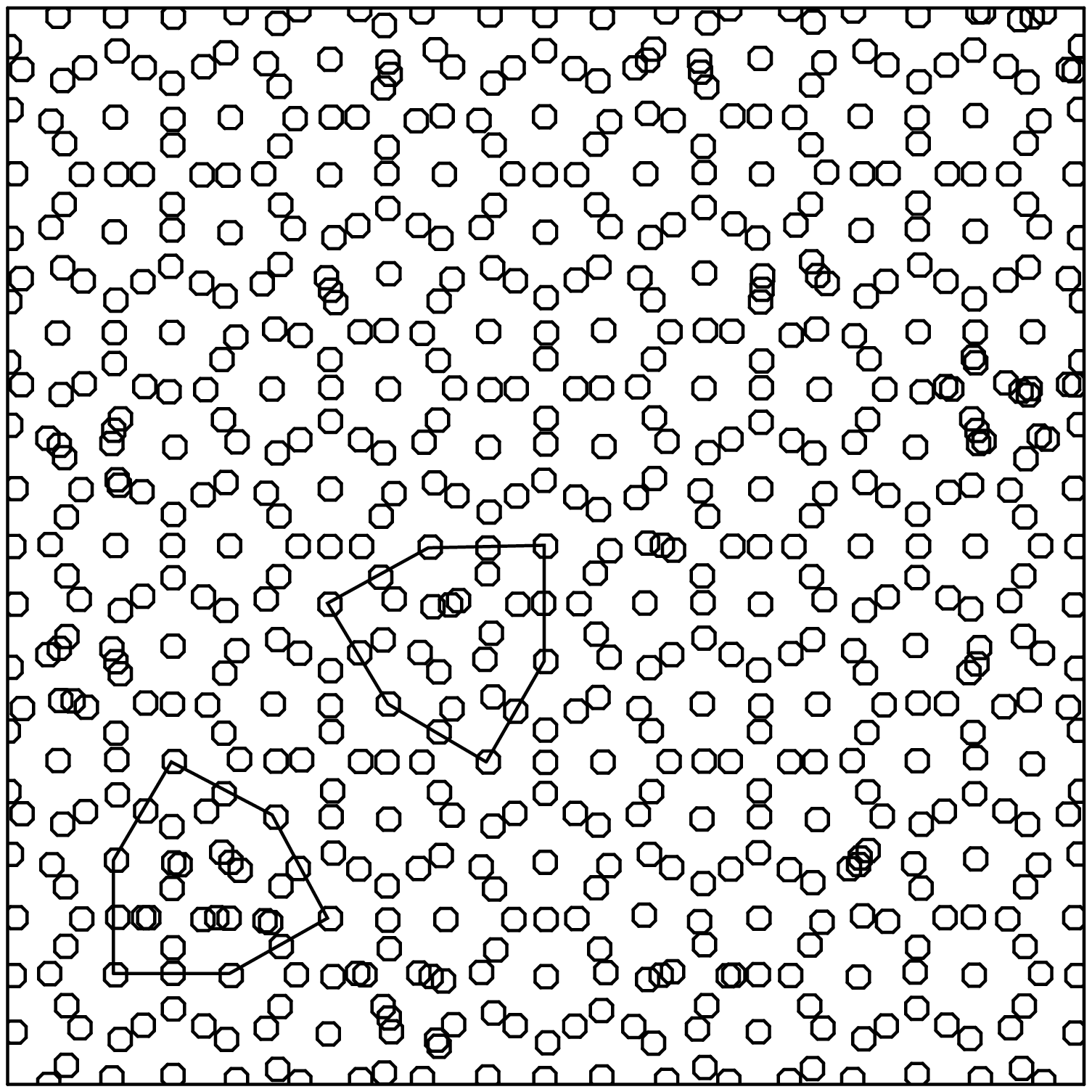}}}
\caption{\small 
  Comparison of the initial configuration with the result after 
  100,000 simulation steps, for a structure based on a tiling with 
  hexagons, containing about 2\% vacancies inside hexagons. 
  On the left, a projection on the xz-plane is shown, with initial
  and final positions connected. It can be seen that atoms primarily
  move vertically, with small horizontal displacements (zig-zags).
  On the right, a projection of the final structure 
  on the xy-plane is shown. The tiling has been reshuffled, and
  at different z-coordinates one has different tilings, so that periodicity
  is slightly broken. Atoms which have moved are primarily inside
  hexagons. Two of the hexagons are indicated.
  }
\label{compconf}
\end{figure}

Our MD simulations were all carried out with a standard 4th-order 
predictor-corrector algorithm\cite{alltil}, at constant temperature and 
constant pressure\cite{lancon} using the constraint method. To keep atoms
sufficiently mobile, a rather high simulation temperature of $T=0.6$ (in
reduced units) was used, compared to a melting temperature of $T_m=0.9$ for 
the idealized structures, and $T_m=0.7$ for Dzugutov's structure,
which was obtained in a cooling simulation \cite{dzu}.
Since Dzugutov's structure 
contains many defects of all kinds, we decided to use, in our 
simulations, the idealized structures described in the last section, 
and to introduce defects in a controlled way where necessary. 
We will start with the description of the results of the cubic case,
where the simulation box contains 10 double layers along the periodic
direction. 
The pure square-triangle tiling structures turned out to be very stable, 
as predicted. The same is true for structures containing asymmetric 
hexagons. Regions with isolated rhombi are found to have a clear 
tendency to transform into hexagons, and then remain immobile. 
Only more disordered structures, where also pairs of rhombi occur, 
are somewhat more mobile.
After very long simulations, these latter 
structures even developed vacancies, whereas the other structures 
only developed half-vacancies. In order to accelerate this process, and 
since Dzugutov's structure also contains many holes, we decided to introduce
some vacancies artificially, in a controlled way, on different classes
of sites. Vacancies are most effective when they are introduced on 
sites in the interior of hexagons, which are involved in the
formation and disappearance of rhombi inside hexagons. When introduced
on other sites, vacancies had a somewhat lesser effect. 
As expected,
mobility is primarily in the periodic direction, whereas
in the quasiperiodic plane atoms move much less (Fig.~\ref{compconf}). It can
clearly be seen, however, that the tiling is reshuffled, and
periodicity is broken.

\begin{figure}
\hspace*{4cm}\includegraphics[width=7.5cm]{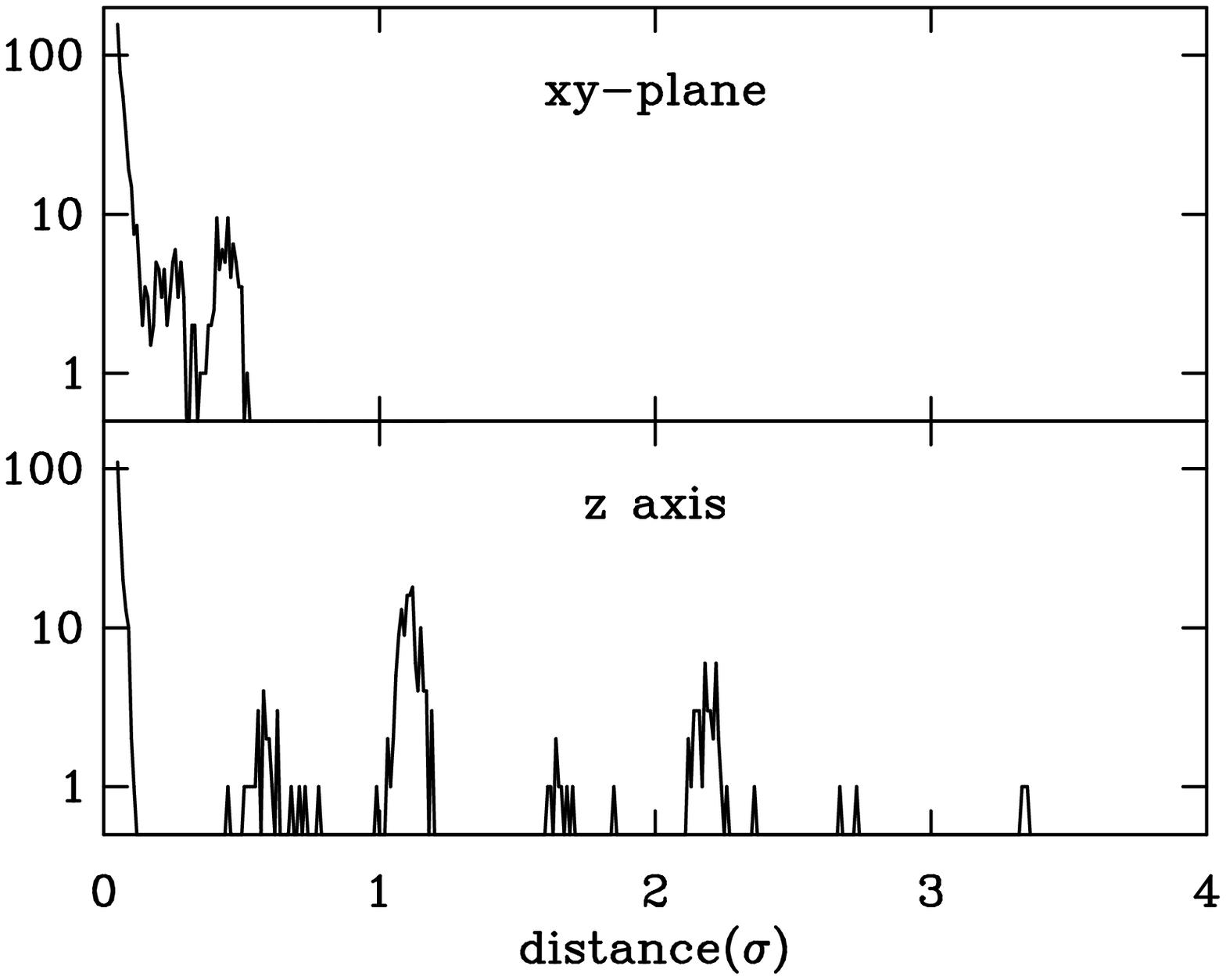}
    \caption{\small
     Histograms of distances between initial and final
     atom positions, after 100,000 simulation steps. Distances
     both in the (periodic) z-direction and within the 
     (quasiperiodic) xy-plane are shown for a structure based 
     on a tiling with hexagons and initial holes.
     In the periodic direction, atoms get much farther, and there
     are clear, discrete step sizes. In the quasiperiodic direction,
     step sizes are much smaller, but there are still discrete step 
     sizes apparent.
  }
\label{histdist}
\end{figure}

In agreement with our theoretical picture, atoms indeed make discrete 
jumps in our simulations (Fig.~\ref{histdist}). These distances
covered are much bigger in the  
periodic direction than in the quasiperiodic direction since in the
first case an atom may jump several times during the simulation.
In the latter case jumps are still discernible, although only
few atoms jump and most don't move at all.

\begin{figure}[hb!]
\centerline{
\parbox{7.2cm}{\includegraphics[width=7.2cm]{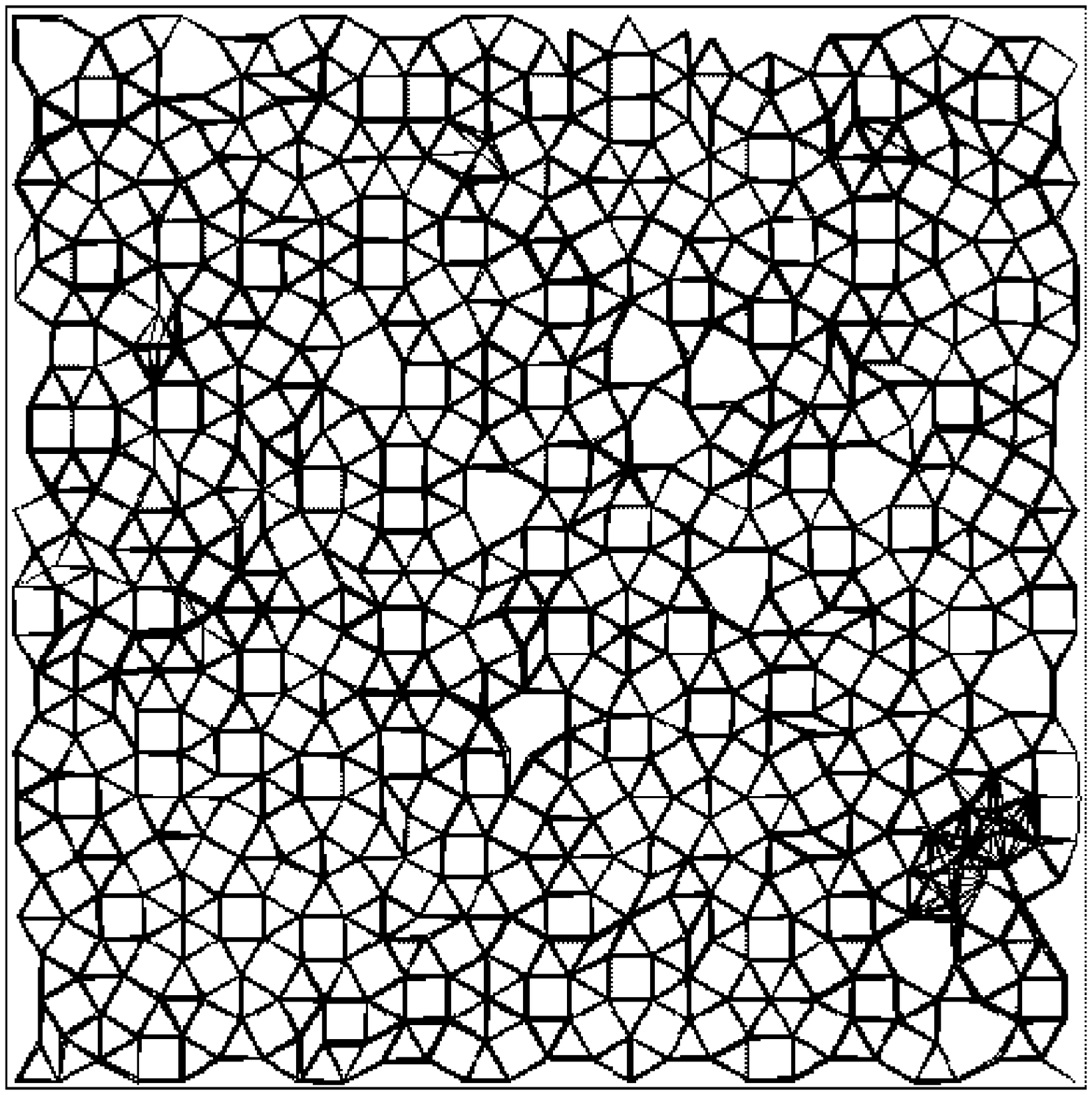}}
\hspace{1cm}
\parbox{7.2cm}{\includegraphics[width=7.2cm]{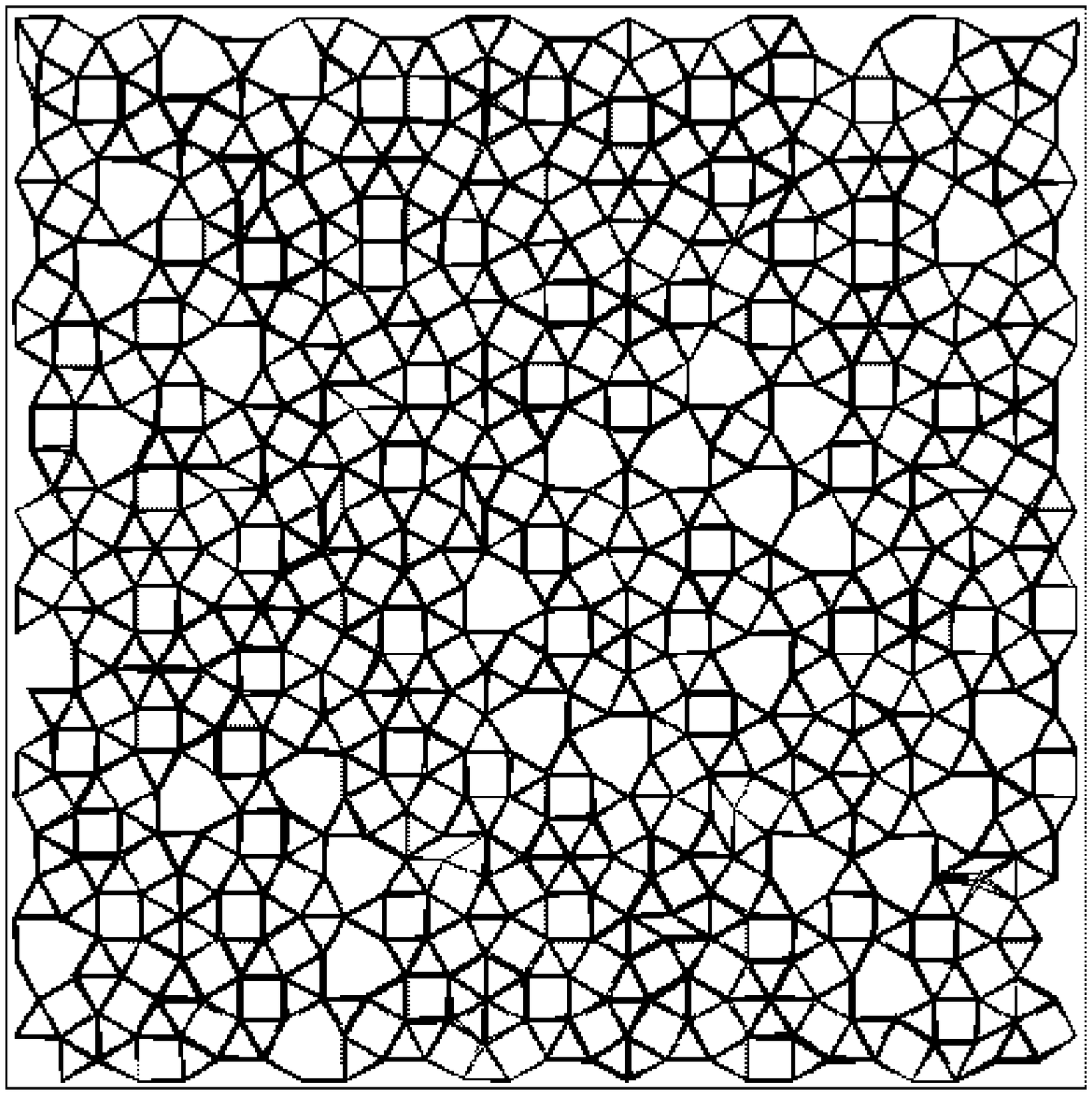}}}
\caption{\small 
  Projection of the initial (left) and final (right) state of the
  tiling after
  700,000 simulation steps for the flat sample. A Comparison shows how
  the tiling has been reshuffled in many places.
  }
\label{flatsim}
\end{figure}

The behaviour of the structural changes is different in the flat tetragonal
sample containing only three double layers along the periodic
direction. Now the flips occur in {\em all} configurations and they do
not terminate, and we
observe an equilibrium structure containing squares, triangles, hexagons and
single rhombi. Even in the pure square-triangle tiling the atoms
become mobile but only if we increase the simulation temperature up to
$T=0.8$. 
Several mobility processes can be distinguished at the tiling level
(Fig.~\ref{flatsim}): 
pairs of rhombi are eliminated immediately. The reason for this will
be discussed in the next section. If there are untilable defects then they
disappear after about 16,000 time steps. The equilibrium density of
hexagons and single rhombi, established by their transition into
one another, is reached after about 70,000 time steps, but the number of
hexagons keeps showing large fluctuation and a very slow relaxation during
the whole simulation run (up to 700,000 time steps).

On the atomic scale we observe phason hopping. This represents the motion of
the atoms corresponding to the tile flip. The hopping is in almost all
cases only local, back and forth. It does not lead to long range
diffusion. Furthermore nearest neighbour jumps typical for ordinary vacancy
diffusion have been found. A few long-range diffusion processes in the
periodic plane have been found, the overwelming part of diffusion
however is along the periodic direction as seen already in the cubic
cell simulations. Therefore the diffusion in this model seems to be
maximally anisotropic.

\section{Activation Energy and Flip Frequency}

For the flat sample it is possible to describe the flipping process
more quantitatively. A jump of an atom chain along the periodic
direction occurs every 110,750 time
steps. The cosine of the angle between the directions of two successive jumps
of the same atom is -0.169 on average, which means  
that successive jumps are nearly uncorrelated.
The frequencies for the jumps are the following: a jump of any tile
occurs every 12,283 time steps, a jump of a rhombus every 21,123 and a jump of
a hexagon every 29,288 steps. For rhombus-hexagon-rhombus the branching ratio
between the backward and the two forward jumps is 4:1:1, and for a
hexagon-rhombus-hexagon jumps it is 13:4.
To determine the activation energies we have optimized the
potential energy of small samples containing the flipable
configurations. The energy of the configuration with the pairs of
rhombi (bottom of Fig.~\ref{basictile}) is about 5.4 (in reduced
units) per period higher than
the energy of the square-triangle configuration. The energy of the hexagon
with the symmetric interior 
is nearly identical to the hexagon containing a rhombus (it is 0.1 per period
lower). If there are three rhombi in the hexagon the energy increases
by 6.4 per period.
Therefore this configuration does not occur.  
The barrier energies can be found by molecular static calculations
\cite{beel}. The jumping atoms are slowly dragged from one
configuration to the other, while monitoring the energy along this
path. In this way it is possible to find the saddle point
configurations and their energies. Special care is needed, however, to
avoid introducing artificial constraints, which might increase the energy.
For the configurations in Fig.~\ref{basictile} it turns out that the
pairs of rhombi actually display saddle point configurations. That is
the reason why 
they disappear immediately at the beginning of the simulation. The barrier for
the flipping process shown in the upper part of Fig.~\ref{basictile} is
too low to be determined accurately, the barrier for the opposite process
(not shown) transforming a hexagon into a second
hexagon via a common central rhombus is also very  
low: 0.5. 
If we take the geometry of the simulation cell into account we find
that there are 
percolating clusters of hexagons sharing a rhombus. Therefore long
range tile diffusion should be possible.

\section{Discussion and Conclusions}

A careful analysis of the final states has shown that flip moves 
indeed occur in our model quasicrystals. Such flip moves have 
also been observed by Dzugutov\cite{dzu2}. 
Their activation energy is prohibitively high if the periodic
direction is long, unless there are other defects present as
well, which can catalyze the flips. 
Half-vacancies and vacancies are 
particularly efficient such catalyzers.
We have found that a sizeable density
of such half-vacancies and vacancies (up to 2\%) is always present in
equilibrium, which therefore makes flip diffusion possible.
Unfortunately, due 
to the (necessary) presence of vacancies and half-vacancies, the 
contributions of flip diffusion and classical vacancy diffusion are 
very hard to separate. We should emphasize, however, that the flip
mechanism, although catalyzed by vacancies, is qualitatively different 
from vacancy diffusion, in that the passage of a vacancy without
flips associated with it leaves the structure unchanged, whereas
with the flips the structure is left completely reshuffled after the
passage of the vacancy. 
If a flat geometry is chosen it is possible to study tiling flips
and their dynamics explicitly. If the starting configuration contains
untilable defects, it is possible to observe tiles move finite
distances. In most of the cases, however, the tiles and the associate
atoms move only back and forth as they do in a
two-well-potential. For geometrical reasons it is not clear why this
should happen, since it is possible for the tiling moves to
percolate. Maybe the long-range diffusion process is too slow to
be seen in molecular 
dynamics. Since we have determined all the energetical parameters for the
flips it should be possible to use them in a Monte Carlo simulation to
clarify these questions and to calculate the diffusion constants
accurately. This is planned for the future.

Our model quasicrystals are somewhat untypical in that they are
periodic in one direction, which causes problems because isolated
flips have to break periodicity and thus lead to larger mismatches.
Moreover, our models are mostly tetrahedrally close-packed, which
makes them very rigid, and which is also not very typical for
quasicrystals.
Still, we believe that our results are relevant
also for (stable) icosahedral quasicrystals, which are completely
non-periodic and not Frank-Kasper like.

\section{Acknowledgements}

The authors would like to thank to M.~Dzugutov for making 
available the structure obtained in his simulation, and C.~Beeli
for suggesting the atomic flip moves shown in Fig.~1.
Part of this work was supported by the Swiss Nationalfonds, Deutsche
Forschungsgemeinschaft, DOE grant DE-FG02-89ER-45405, and
Packard Foundation grant FDN 89-1606 during a common stay of the
authors at LASSP, Cornell University, Ithaca, New York.
The computations were performed on facilities of Cornell Materials
Science Center and of the Institut f\"ur Theoretische und Angewandte Physik.

\end{document}